\documentclass[aps,pra,onecolumn,groupedaddress]{revtex4}

\usepackage{graphics}
\usepackage{epsfig}

\begin{document}

\title{A Quantum Teleportation Game}
\author{Stefano Pirandola}

\affiliation{{\it Dipartimento di Fisica, Universit\`a di
Camerino, I-62032 Camerino, Italy \\ stefano.pirandola@unicam.it}}

\begin{abstract}
We investigate a game where a sender (Alice) teleports coherent
states to two receivers (Bob and Charlie) through a tripartite
Gaussian state. The aim of the receivers is to optimize their
teleportation fidelities by means of local operations and
classical communications. We show that a non-cooperative strategy
corresponding to the standard telecloning protocol can be
outperformed by a cooperative strategy which gives rise to a novel
(cooperative)\ telecloning protocol.
\end{abstract}

\maketitle
\section{Introduction}

The theory of games has recently entered the domain of quantum
mechanics \cite{game1}, giving rise to the so-called ``quantum
games'', a new\ way of looking at concepts of quantum information
and computation. Beside the advantages connected to novel quantum
strategies, it is possible to consider a set of classical
strategies to be optimized over a shared quantum signal. This is
the case of the present paper, where the interpretation of a
quantum teleportation network as a game played by more parties
leads to the definition of a new (cooperative) telecloning
protocol which can outperform the standard (non-cooperative)\ one.

\section{The Game}

\begin{figure}[ptbh]
\vspace{0.0cm}
\par
\begin{center}
\includegraphics[width=0.6\textwidth]{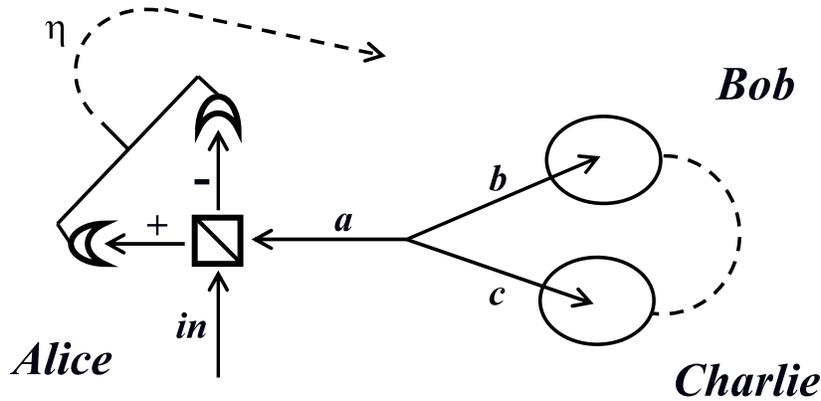}
\end{center}
\par
\vspace{0.0cm}\caption{Set-up of the game. Circles are
local operations and dashed lines are classical communications.}%
\label{setup}%
\end{figure}

In our game, Alice has a large set of coherent states $\{\left|
\varphi\right\rangle _{in}\}$ which she wants to teleport to both
Bob and Charlie. For every input $\left|  \varphi\right\rangle
_{in}$, she can exploit a shared quantum channel given by a 3-mode
Gaussian state $\rho_{abc}$\ with modes $a,b,c$ belonging to
Alice, Bob and Charlie respectively (see Fig.\ref{setup}). In
order to make the game fair, such state must be symmetric for Bob
and Charlie, i.e. $\rho_{abc}=\rho_{acb}$. Following the standard continuous variable teleportation protocol \cite{telepo}%
, Alice mixes her part of the channel\ with the input state
through a balanced beam-splitter and performs an homodyne
detection of the output modes, i.e. she measures the quadratures
$\hat{X}_{-}\equiv2^{-1/2}(\hat{X}_{a}-\hat{X}_{in})$ and
$\hat{P}_{+}\equiv2^{-1/2}(\hat{P}_{a}+\hat{P}_{in})$. After this
\textit{Bell measurement}, Alice announces the result $\eta
\equiv-X_{-}+iP_{+}$ over a public classical channel so that both
Bob and Charlie can accomplish teleportation by a set of local
operations and classical communications (Fig.\ref{setup}).
The aim of Bob and Charlie is to maximize the fidelity
$F_{AB},F_{AC}$ of the copy teleported to their respective modes
$b$ and $c$ over a large number of instances of the game.

The study of this problem is hard in general, but, if we restrict
to a particular set of shared channels, we are already able to
show remarkable effects of cooperation. For simplicity consider a
zero-drift channel characterized by a correlation
matrix (CM) of the form%
\begin{equation}
V=\left(
\begin{array}
[c]{ccc}%
\alpha I & \delta Z & \delta Z\\
\delta Z & \beta I & \gamma I\\
\delta Z & \gamma I & \beta I
\end{array}
\right)  \label{CM}%
\end{equation}
where $I$ is the $2\times2$ identity matrix, $Z\equiv diag(1,-1)$
the $\sigma_{z}$\ Pauli matrix and $\alpha,\beta,\gamma,\delta$
real parameters. Such form is symmetric under the exchange
Bob$\longleftrightarrow$Charlie and, tracing out one mode of the
three, it implies bipartite entanglement only between the sender
(Alice) and the receiver (Bob or Charlie). The CM\ (\ref{CM}) is
genuine (i.e. it corresponds to a physical state \cite{genuine})
if and only if the uncertainty principle $V-J/2\geq0$
holds, where $J\equiv\oplus_{k=a,b,c}Y_{k}$ with $Y_{k}$ the $\sigma_{y}%
$\ Pauli matrix. We can achieve this condition if we choose
\begin{equation}
\alpha\geq\frac{1}{2},\quad\beta=\frac{1}{2}(\alpha+1),\quad\gamma
=\frac{\alpha}{2},\quad\delta=\frac{1}{2}\sqrt{(2\alpha-1)(\alpha+1)}
\label{rel}%
\end{equation}
so that we simply deal with a one-variable CM $V=V(\alpha)$.

\section{Non-Cooperative Strategy}

\begin{figure}[ptbh]
\vspace{-0.0cm}
\par
\begin{center}
\includegraphics[width=0.45\textwidth]{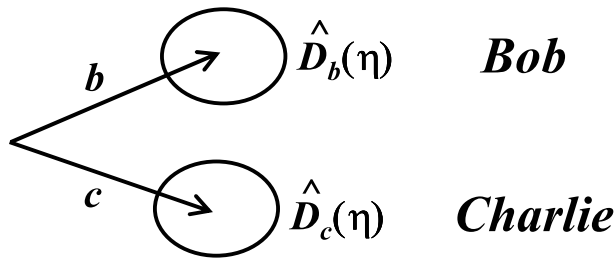}
\end{center}
\par
\vspace{-0.0cm}\caption{Non-cooperative strategy.}%
\label{NC}%
\end{figure}

A non-cooperative strategy for this game is simply given by the
standard telecloning protocol \cite{teleclo}. In this strategy,
Bob and Charlie ignore each other and apply suitable conditional
displacements on their modes which optimize their\ teleportation
fidelities $F_{AB}^{tr}$ and $F_{AC}^{tr}$. Such conditional
displacements are $\hat{D}_{b}(\eta)=\exp(\eta\hat
{b}^{\dagger}-\eta^{\ast}\hat{b})$\ and$\ \hat{D}_{c}(\eta)$,
where the shift $\eta$\ exactly balances the drift created by
Alice's measurement on modes $b$ and $c$ (see Fig.\ref{NC}).
The symmetry of the channel implies $F_{AB}^{tr}=F_{AC}^{tr}\equiv F^{tr}$ and
$F^{tr}$ is simply related to the CM of the reduced channel $\rho^{tr}%
=Tr_{b}(\rho_{abc})=Tr_{c}(\rho_{abc})$.

Quantitatively we have \cite{fiura}
$F^{tr}=(\det\Gamma^{tr})^{-1/2}$ where
$\Gamma^{tr}=(1+\alpha+\beta-2\delta)I$. Setting
$\kappa(\alpha)\equiv(1+\alpha+\beta-2\delta)$, we simply have
$\Gamma^{tr}=\kappa(\alpha)I$ and therefore$\
F^{tr}=\kappa(\alpha)^{-1}$.

\section{Cooperative Strategy}

\begin{figure}[ptbh]
\vspace{-0.0cm}
\par
\begin{center}
\includegraphics[width=0.6\textwidth]{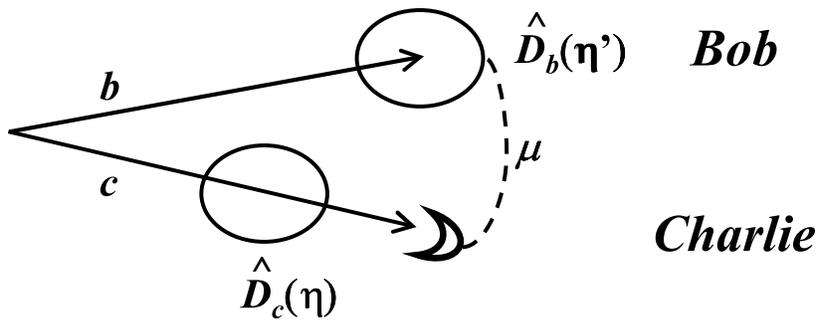}
\end{center}
\par
\vspace{-0.0cm}\caption{Single instance of the cooperative
strategy.  Charlie displaces his mode, heterodynes it and
communicates the result to Bob, who performs a modified
displacement.}%
\label{C}%
\end{figure}

A simple cooperative strategy for the two receivers can be derived
if we include measurements in the set of local operations. Suppose
that, after Alice's declaration of measurement result $\eta$,
Charlie performs the usual conditional displacement
$\hat{D}_{c}(\eta)$, but then, he heterodynes his mode and
reconstructs the state from the result $\mu$. This subsequent
operation surely worsens his fidelity ($F_{AC}<F_{AC}^{tr}$) but,
if he now communicates the result to Bob, then Bob can actually
improve his teleportation ($F_{AB}\geq F_{AB}^{tr}$). It is
sufficient that Bob performs a conditional displacement
$\hat{D}_{b}(\eta^{\prime})$ where
$\eta^{\prime}=\eta+(\beta+1/2)^{-1}(\delta -\gamma)(\mu-\eta)$ is
a modified shift which balances both the drift caused by Alice's
measurement and the drift acquired from the Charlie's local
operations \cite{network} (see Fig.\ref{C}).

Quantitatively we have for Charlie
$F_{AC}=(\det\Gamma_{AC})^{-1/2}$ where
$\Gamma_{AC}=I+\Gamma^{tr}$. It is easy to prove that
$F_{AC}<F_{AC}^{tr}$ and $F_{AC}\leq1/2$ (threshold for quantum
teleportation \cite{fuchs}), and introducing $\kappa(\alpha)$, as
before, we have $F_{AC}=[\kappa (\alpha)+1]^{-1}$. Meanwhile, for
Bob, it is possible to prove that
\begin{equation}
F_{AB}=\frac{\alpha+2}{(\alpha+2)\kappa(\alpha)-2(\delta-\gamma)^{2}}
\label{F_AB}
\end{equation}
which clearly satisfies $F_{AB}\geq F_{AB}^{tr}$ for every $\alpha\geq1/2$.

Due to the symmetry of the channel, the performances are exactly
inverted if the roles of Bob and Charlie are inverted (i.e. Bob
displaces, measures and informs Charlie, who performs a modified
displacement). Obviously in the case where always the same
receiver performs the measurement, the strategy is lossy for that
receiver while it is highly convenient for the other one. However
a fair cooperation between the receivers is possible alternating
the roles during the various instances of the game (or choosing
the roles by a truly random generator). In such a case both
receivers achieve the same fidelity $F=(F_{AB}+F_{AC})/2$ which
can be greater than $F^{tr}$.

\begin{figure}[ptbh]
\vspace{-0.0cm}
\par
\begin{center}
\includegraphics[width=0.5\textwidth]{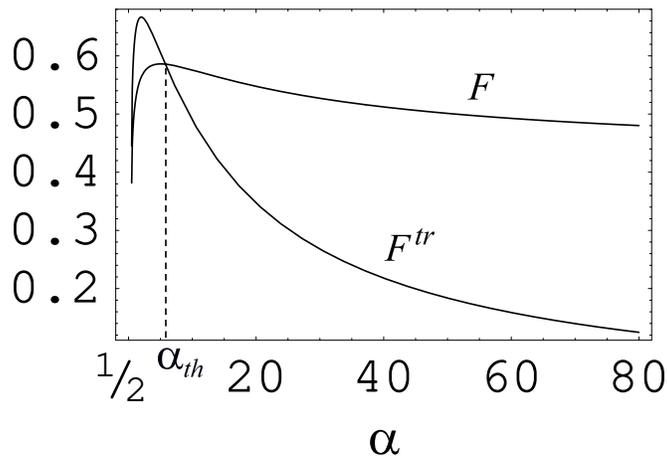}
\end{center}
\par
\vspace{-0.0cm}\caption{Fidelities $F$ and $F^{tr}$ versus noise
parameter $\alpha$.} \label{graf}
\end{figure}

The two fidelities $F$ and $F^{tr}$ as functions of parameter
$\alpha$\ are plotted in Fig.\ref{graf}, where we can see a
threshold value $\alpha _{th}\sim5.76$.
For $\alpha>\alpha_{th}$, i.e. for a noisy channel, the
cooperative strategy ($F$)\ is better, and it remains above the
classical teleportation value ($1/2$) up to high values of
$\alpha$. The non-cooperative strategy ($F^{tr}$) is instead
better for $\alpha <\alpha_{th}$ where the channel is more pure
and where it is possible to reach the maximum fidelities permitted
by the no-cloning theorem ($F^{tr}=2/3$ for $\alpha=2$). The
depicted situation clearly shows that a cooperative telecloning
protocol, as naturally defined by the cooperative strategy, can be
more efficient of the standard one in presence of noisy channels.

A final remark has to be made about the measurement. The choice of
the heterodyne detection, as local measurement for the receivers
(see Fig.\ref{C}), is strictly connected with the particular form
(\ref{CM}) of the CM, since it just represents an optimal local
measurement in order to maximize teleportation fidelity in that
case \cite{POVM}. If we consider a Gaussian channel with arbitrary
CM (in general asymmetric for the receivers), then an optimal
local measurement (in general different for the two receivers) is
equivalent to a suitable squeezing transformation followed by an
heterodyne detection \cite{POVM}. For this reason the cooperative
protocol can be extended to more general channels if we suitably
modify the local measurement performed by the receivers.

\section{Conclusion}

A simple teleportation game between a sender and two receivers has
been investigated. A first non-cooperative strategy is given by
the standard telecloning protocol, but this one can be
outperformed by a cooperative strategy which represents a new
(cooperative) telecloning protocol. This is surely true for
channels of the type (\ref{CM}) with a certain amount of noise
($\alpha>\alpha_{th}$), but more general channel could be
considered.

\end{document}